\documentclass[%
reprint,
superscriptaddress,
frontmatterverbose,
showpacs,
preprintnumbers,
nofootinbib,
nobibnotes,
amsmath,amssymb,
aps,
 prl,
]{revtex4-2}
\usepackage[pdftex]{graphicx}
\usepackage[caption=false]{subfig}
\usepackage{amsmath}
\usepackage{mathtools}
\usepackage{dcolumn}
\usepackage{bm}
\usepackage[pdftex]{color}
\usepackage{CJK}
\usepackage[T1]{fontenc}
\usepackage{float}
\usepackage{mathrsfs}
\usepackage{mathbbol}
\usepackage{hyperref}
\hypersetup{
colorlinks = true,
linkcolor = [rgb]{0.70,0.13,0.13},
citecolor = [rgb]{0.13,0.55,0.13},
urlcolor = [rgb]{0.25, 0.41, 0.88}}

\def\ve#1{{\bm{#1}}}
\def\e{\mathrm{e}}
\def\ii{\mathrm{i}}

\let\temp\epsilon
\let\epsilon\varepsilon
\let\varepsilon\temp
\let\temp\relax
\let\temp\phi
\let\phi\varphi
\let\varphi\temp
\let\temp\relax
\begin{document}
%
\begin{CJK*}{UTF8}{}
  \title{Machine Learning for Ground State Preparation via Measurement and Feedback}
  \author{Chuanxin Wang (\CJKfamily{gbsn}{王传新})}
  \affiliation{
    College of Physics,
    Jilin University,
    Changchun 130012, China}
  \affiliation{
    Department of Physics,
    University of California,
    San Diego, La Jolla, CA 92037, USA}
  \author{Yi-Zhuang You (\CJKfamily{gbsn}{尤亦庄})}
  \affiliation{
    Department of Physics,
    University of California,
    San Diego, La Jolla, CA 92037, USA}
  %
  \begin{abstract}
  We present a recurrent neural network-based approach for ground state preparation utilizing mid-circuit measurement and feedback. 
  Unlike previous methods that use machine learning solely as an optimizer, our approach dynamically adjusts quantum circuits based on real-time measurement outcomes and learns distinct preparation protocols for different Hamiltonians. 
  Notably, our machine learning algorithm consistently identifies a state preparation strategy wherein all initial states are first steered toward an intermediate state before transitioning to the target ground state. We demonstrate that performance systematically improves as a larger fraction of ancilla qubits are utilized for measurement and feedback, highlighting the efficacy of mid-circuit measurements in state preparation tasks.
  \end{abstract}
  \maketitle
\end{CJK*}
%
\section{Introduction}
\label{sec:level1}
Ground state preparation is a fundamental task in quantum information science.
The variational quantum eigensolver (VQE) is a hybrid quantum-classical algorithm that leverages the variational principle to prepare an approximate ground state of a given Hamiltonian \cite{Peruzzo2014-vp}.
Traditionally, VQE implementations rely on unitary operations to prepare the ground state
\cite{PhysRevLett.122.140504,
grimsley2019adaptive,
PhysRevResearch.6.013254,
zhang2022variational,
mihalikova2022cost,
Chivilikhin2020MoGVQEMG,
PhysRevLett.122.230401,
doi:10.1021/acs.jctc.8b00943,
cerezo2022variational,
kirby2021contextual,
yordanov2021qubit}.
Recent advances have demonstrated that mid-circuit measurements with feedback can reduce computational resources in quantum information processing \cite{PRXQuantum.3.040337,
PhysRevX.14.021040,
PhysRevLett.127.220503,
PRXQuantum.4.020315,
Malz_2024,
yan2024variationalloccassistedquantumcircuits}.
These approaches involve performing measurements during the execution of unitary operations, with subsequent operations conditioned on the measurement outcomes.
Designing an effective quantum protocol that integrates measurement-based feedback to control unitary gates remains an open challenge for ground-state preparation tasks.
\par
Machine learning (ML) has recently attracted interest as a method for controlling unitary gates conditioned on measurement outcomes \cite{alam2024learningdynamicquantumcircuits,
puente2025learningfeedbackmechanismsmeasurementbased}.
In the context of optimizing VQE \cite{
NEURIPS2021_97244127,
PhysRevB.102.075104,
patel2024curriculumreinforcementlearningquantum,
kundu2024kanqas,
kundu2025improvingthermalstatepreparation}, most prior studies have employed ML as an optimization tool, solving instances on a case-by-case basis.
In contrast, we aim to develop a protocol in which ML learns to dynamically control quantum circuits based on measurement feedback. 
Furthermore, our algorithm learns to generate different state preparation protocols conditioned on  different target Hamiltonians as input information.
\par
In this paper, we propose a recurrent neural network (RNN) framework for ground-state preparation. 
The ground state is represented as a density matrix that minimizes the energy expectation value of the Hamiltonian, which serves as the loss function for RNN optimization.
The initial state is an arbitrary pure or mixed state, unknown to the RNN. 
The preparation process unfolds over multiple time steps, with measurements performed at the end of each step. 
After each unitary operation and measurement cycle, the RNN updates its control strategy based on all prior measurement outcomes, adjusting subsequent quantum gates accordingly. 
Additionally, the RNN integrates measurement outcomes to infer the current state and prepare the target ground state based on its knowledge about the Hamiltonian.
\par
This paper is organized as follows: 
In Sec.~\ref{Sect:III}, we illustrate the design of the RNN and the training strategy.
The results of single- and two-qubit state preparation are discussed in Sec.~\ref{Sect:IV}.
Finally, the summary and future perspectives are presented in Sec.~\ref{Sect:V}.
\section{Method}
\label{Sect:III}
\subsection{Initial State Setting}
\label{sec:level2,1}
For a system with $N$ qubits, we define $N^\text{sys}$ as the number of system qubits and $N^{\text{anc}}$ as the number of ancilla qubits, such that $N = N^{\text{sys}} + N^{\text{anc}}$.
\par
\begin{figure}[tb]
    \centering
\includegraphics[scale=0.6]{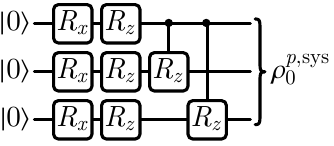} 
    \caption{The architecture of the quantum circuit for the 3-qubit pure-state preparation.
    }
    \label{initial_circuit}
\end{figure}

To be general, we assume that the initial state of the system qubits can be either pure or mixed. It can also be sampled randomly. For example, Fig.~\ref{initial_circuit} illustrates a quantum circuit used to prepare a random pure initial state, $\rho_0^{p,\text{sys}}$.
Single-qubit rotation gates, $R_x=\e^{\ii(\theta/2)\sigma_x}$ and $R_z=\e^{\ii(\phi/2)\sigma_z}$, are applied with gate angles $\theta,\phi$ randomized in the range $\left[0,2\pi\right]$, allowing the initial $|0\rangle$ state to be mapped to any point on the Bloch sphere.
Entanglement among qubits is introduced through controlled $R_z$ gates.
\par
The mixed initial state of the system qubits is constructed by sampling a statistical mixture of random pure states:
\begin{equation}
    \rho_0^{m,\text{sys}} = \sum_{q} p_{q} \rho_{0,q}^{p,\text{sys}},
\end{equation}
where $\rho_{0,q}^{p,\text{sys}}$ is the $q$-th pure state generated by the circuit in Fig.~\ref{initial_circuit}.
The coefficients $p_q$ are randomly sampled within $\left(0,1\right)$ and satisfy the normalization condition $\sum_{q} p_{q} = 1$.
\par
The initial state of the ancilla qubits at each time step $t$ is given by
\begin{equation}
    \rho_{t}^{\text{anc}} = \left(|0\rangle \langle0|\right)^{\otimes N^{\text{anc}}}.
\end{equation}
The total initial state of the system and ancilla qubits is then
\begin{equation}
    \rho_{0} = \rho_0^{\text{sys}} \otimes \rho_0^{\text{anc}},
\end{equation}
where $\rho_0^{\text{sys}}$ can either a random pure state $\rho_0^{p,\text{sys}}$ or a random mixed state $\rho_0^{m,\text{sys}}$ in our implementation. In this way, our algorithm can be trained to prepare target ground states from various generic initial states.
\par
\par
\subsection{Quantum Circuits and Measurements}
\label{sec:level2,2}
\par
\begin{figure}[tb]
    \centering
\includegraphics[scale=0.6]{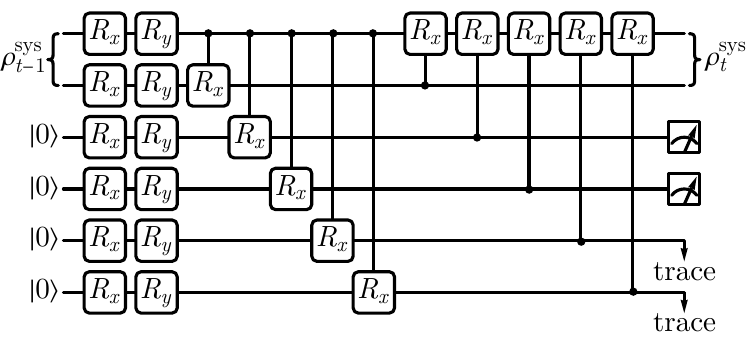} 
    \caption{The architecture of the quantum circuit $\text{QC}_{i}$ in the case of two-qubit state preparation.
    }
    \label{circuit}
\end{figure}

The quantum circuits are implemented using \textsc{PennyLane} \cite{bergholm2018pennylane}.
For a $T$ time step preparation, each quantum circuit consists of the same type of unitary operations $U_{\theta_{t}}$ 
From $t = 0$ to $t = T-1$, measurements $P_{\ve{m}_{t}}$ are performed, while no measurement is conducted in the final step.
The quantum circuit $\text{QC}_t$ for a two-qubit preparation is shown in Fig.~\ref{circuit}.
It consists of $N^{\text{sys}} = 2$ system qubits, $N^{\text{anc}}_{m} = 2$ ancilla qubits for measurement and feedback, and $N^{\text{anc}}_{t} = 2$ ancilla qubits that are traced out.
The unitary operator $U_{\theta_{t}}$ is parametrized by gate parameters $\theta_t$, which define single-qubit rotations around the $X$ and $Y$ axes, as well as controlled-$X$ gates that introduce entanglement.
\par
In each quantum circuit $\text{QC}_t$, the input state is given
\begin{equation}
    \rho_{t-1} = \rho_{t-1}^{\text{sys}} \otimes \rho_{t-1}^{\text{anc}}.
\end{equation}
The unitary operator $U_{\theta_{t}}$ and projector $P_{\ve{m}_{t}}$ are then applied to the input state to compute the output state $\rho_{t}^{\text{sys}}$ before tracing out ancilla qubits:
\begin{equation}
\begin{split}
    \rho_{t}^{\text{sys}} &= \mathop{\mathrm{Tr}}\left( P_{\ve{m}_{t}} U_{\theta_{t}} \rho_{t-1} U_{\theta_{t}}^{\dagger} P_{\ve{m}_{t}}\right)\\
    &:= 
    \label{eq7}\mathcal{C}_{\theta_{t},\ve{m}_{t}}\left[\rho_{t-1}^{\text{sys}} \right],
\end{split}
\end{equation}
which defines the quantum process $\mathcal{C}_{\theta_{t},\ve{m}_{t}}$ at step $t$. The projector $P_{\ve{m}_{t}}$ reads
\begin{equation}
    P_{\ve{m}_{t}} = \prod_{l} \frac{\mathbb{1}+m_{t,l}\sigma_z}{2},
\end{equation}
where $\ve{m}_t$ labels the binary string of measurement outcomes at step $t$
\begin{equation}
    \ve{m}_t = \left(m_{t,1},m_{t,2}, ...,m_{t,N^{\text{anc}}_{m}}\right) \in  \left\{0, 1\right\}^{\otimes N^{\text{anc}}_{m}},
\end{equation}
with $m_{t,l}$ being the measurement outcome of the $i$-th qubit.
$\mathbb{1}$ is the identity operator, and $\sigma_z$ is the Pauli-Z operator. 
\par
\par
%
\subsection{RNN Architecture}
\label{sec:level2,3}
\par
\begin{figure}[tb]
    \centering
\includegraphics[width=1.0\linewidth]{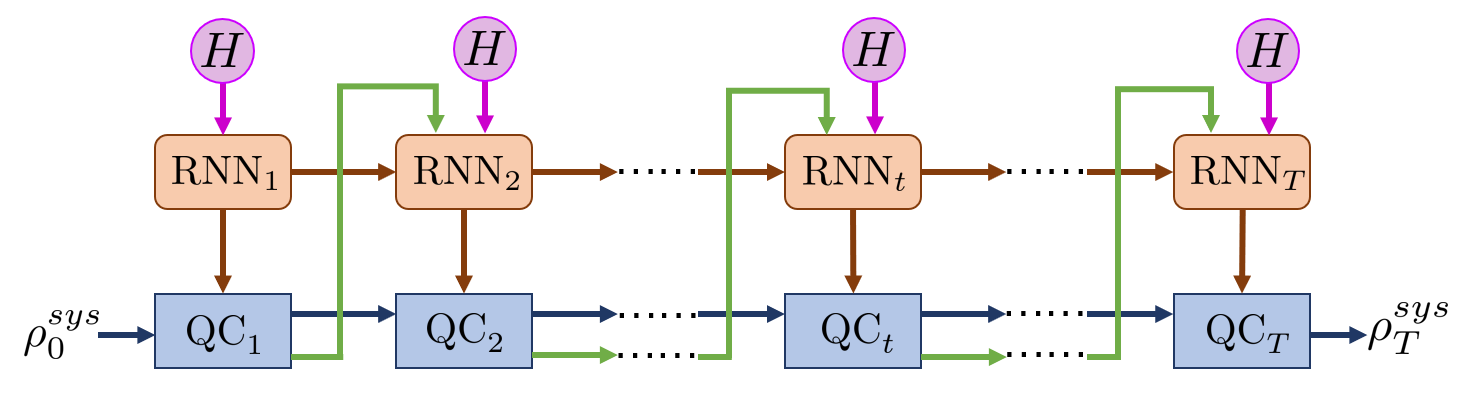} 
    \caption{Structure of an $n$-step RNN controlled quantum state preparation. Red arrows: classical control/memory information flow. Green arrows: measurement outcome feeding. Blue arrows: quantum state evolution. Pink arrows: injections of Hamiltonian information.
    }
    \label{RNN}
\end{figure}

The architecture of the RNN is shown in Fig.~\ref{RNN}, which is implemented using the \textsc{TensorFlow} platform \cite{Tensorflow}.
The brown squares represent LSTM-type RNN cells \cite{hochreiter1997long}, while the blue squares correspond to quantum circuits.
At each time step $t$, the RNN cell takes all previous measurement outcomes and a given Hamiltonian as inputs to compute the gate parameters $\theta_t$:
\begin{equation}
\label{eq10}
    \theta_t = \mathcal{F}_{\phi}\left( \ve{m}_1,\ve{m}_2, ...,\ve{m}_{t-1},H \right),
\end{equation}
where $\mathcal{F}$ denotes the LSTM cell computation, and subscript $\phi$ indicates RNN parameters.
The output of the RNN, $\theta_t$, serves as the parameters for the unitary operator $U_{\theta_t}$, with brown arrows indicating the information flow between RNN cells and quantum circuits.
The pink and green arrows represent the input of the Hamiltonian $H$ and the passing of the measurement outcomes $\ve{m}_{t-1}$, respectively.
Measurement outcomes prior to time step $t$ are effectively stored in the memory cell states, which are passed down along the brown arrows from one RNN cell to the next, along with the hidden states.
\par
\subsection{Loss Function}
\label{sec:level2,4}
\par
Given the Hamiltonian $H$ as input, starting from the initial state $\rho_0^\text{sys}$, Eqs.~\eqref{eq7} and \eqref{eq10} jointly specify a quantum-classical recurrent dynamics that prepares $\rho_T^\text{sys}(\phi)$ after $T$ steps, which is ultimately parametrized by the RNN model parameter $\phi$, as illustrated in Fig.~\ref{RNN}.

The loss function can be defined as the expectation value of the energy of the given Hamiltonian $H$ on the generated state $\rho_T^\text{sys}$,
\begin{equation}
\label{eq11}
    E(\phi) = \mathop{\mathrm{Tr}}\left( \rho_{T}^{\text{sys}}(\phi) H\right).
\end{equation}
Since the ground state energy $E_{\text{min}}$ corresponds to the minimum of the energy spectrum, minimizing Eq.~\eqref{eq11} yields both an estimation of  $E_{\text{min}}\approx \min_\phi E(\phi)$, and the corresponding protocol to prepare the ground state $\rho_{T}^{\text{min}}$ as the RNN gets trained.
\par
In terms of the Hamiltonian used to train the algorithm, for single-qubit preparation, the randomly sampled Hamiltonian is given by
\begin{equation}
\label{eq3}
    H = h_x \sigma_x + h_y \sigma_y + h_z \sigma_z,
\end{equation}
where $\sigma_x$, $\sigma_y$, $\sigma_z$ are Pauli matrices.
The coefficients $h_x$, $h_y$, $h_z$ are randomly sampled in $\left[-1,1\right]$.
For an $n$-qubit system, the general form of the Hamiltonian is a linear combination of all Pauli string operators as
\begin{align}
    H = & \sum_{a,b, ..., c = x,y,z,I} h_{ab ... c} \sigma_{a}^{1}\otimes\sigma_{b}^{2}\otimes ...\otimes \sigma_{c}^{n},
\end{align}
where $\sigma_I$ is the identity operator, and the superscripts denote the qubit labels.
The real coefficients $h_{ab ... c}$ are sampled within $\left[-1,1\right]$ uniformly. We expect the RNN to learn the ground state preparation strategy for any generic Hamiltonian after training, although this method is not scalable for large systems. We suspect that imposing locality constraints on the Hamiltonian and training on quantum devices may enable the algorithm to gain scalability, but we will leave that to future research.
\subsection{Training Strategy}
\label{sec:level2,5}
For each epoch during training, Eq.~\eqref{eq7} is repeated $T$ times to calculate the output state of system qubits $\rho_T^{\text{sys}}$, where Eq.~\eqref{eq10} generates parameters $\theta_t$ of unitary gates.
The loss function is then computed using the output $\rho_T^{\text{sys}}$ from Eq.~\eqref{eq11}.
The RNN parameters $\theta_t$ are optimized using the \textsc{Adam} optimizer \cite{DBLP:journals/corr/KingmaB14} through backpropagation.
The initial state of system qubits $\rho_0^{\text{sys}}$, and the Hamiltonian $H$ are randomly sampled as described in Subsections \ref{sec:level2,1} and \ref{sec:level2,4}.
\section{Results}
\label{Sect:IV}
\subsection{Single-qubit State Preparation}
\label{sec:level3,1}
In the case of single-qubit preparation, we take $n=5$ time steps, using two ancilla qubits for measurements and two for tracing out.
The initial states are mixed states.
The randomly sampled Hamiltonian follows the general construction in Eq.~\eqref{eq3}.
\par
\begin{figure*}[tb]
\begin{minipage}{.166\textwidth}
  \centering
  \includegraphics[width=1.3\linewidth]{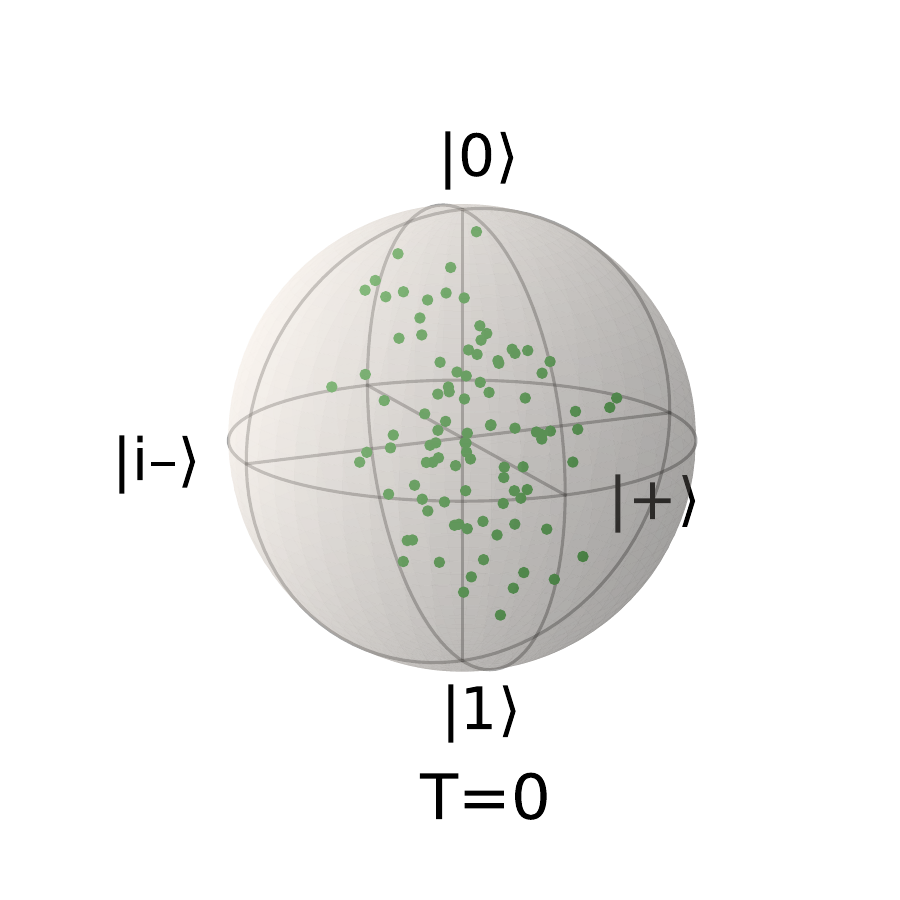}
\end{minipage}%
\begin{minipage}{.166\textwidth}
  \centering
  \includegraphics[width=1.3\linewidth]{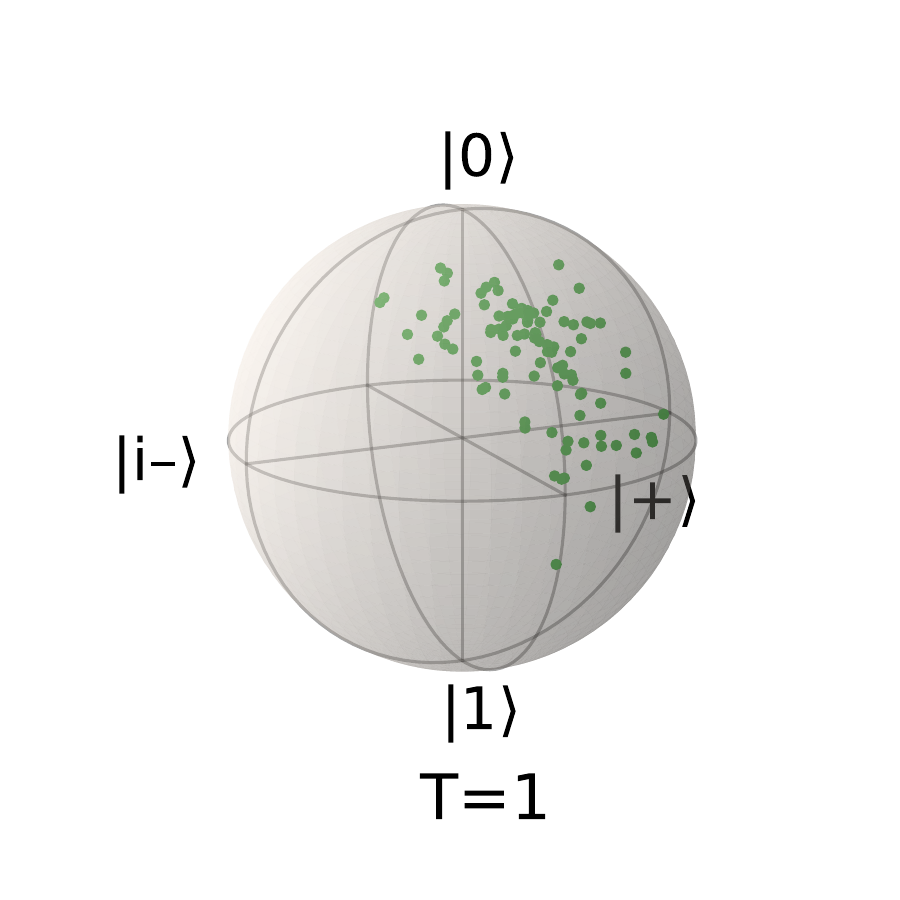}
\end{minipage}%
\begin{minipage}{.166\textwidth}
  \centering
  \includegraphics[width=1.3\linewidth]{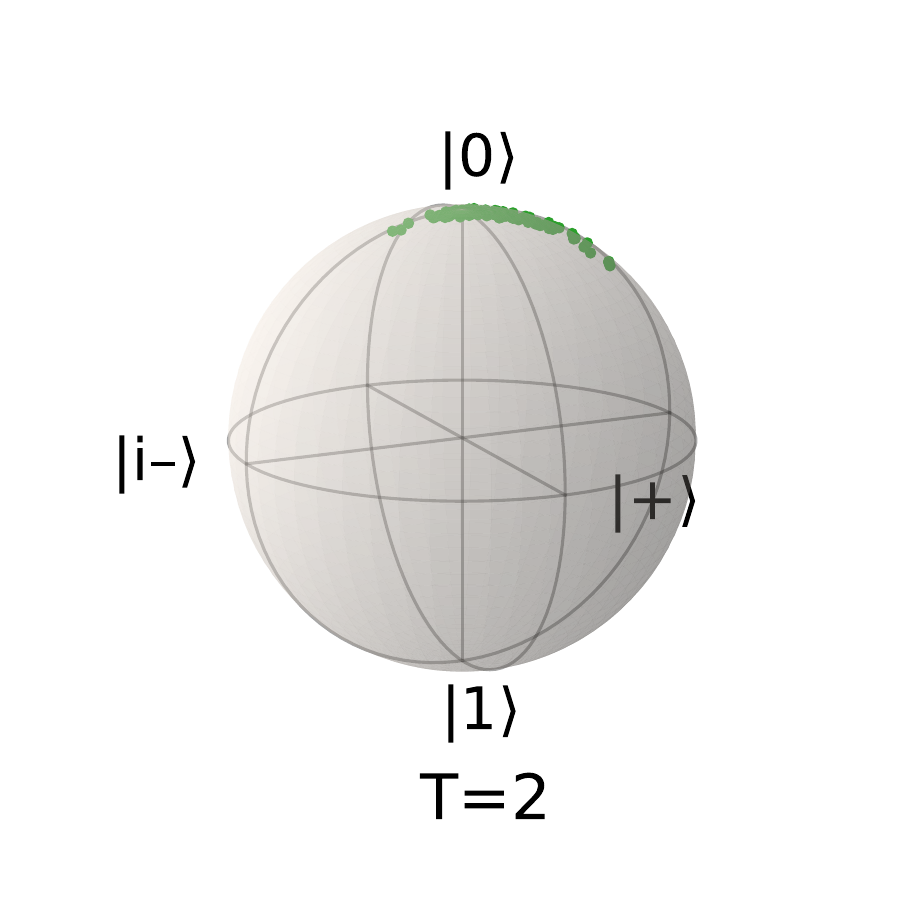}
\end{minipage}%
\begin{minipage}{.166\textwidth}
  \centering
  \includegraphics[width=1.3\linewidth]{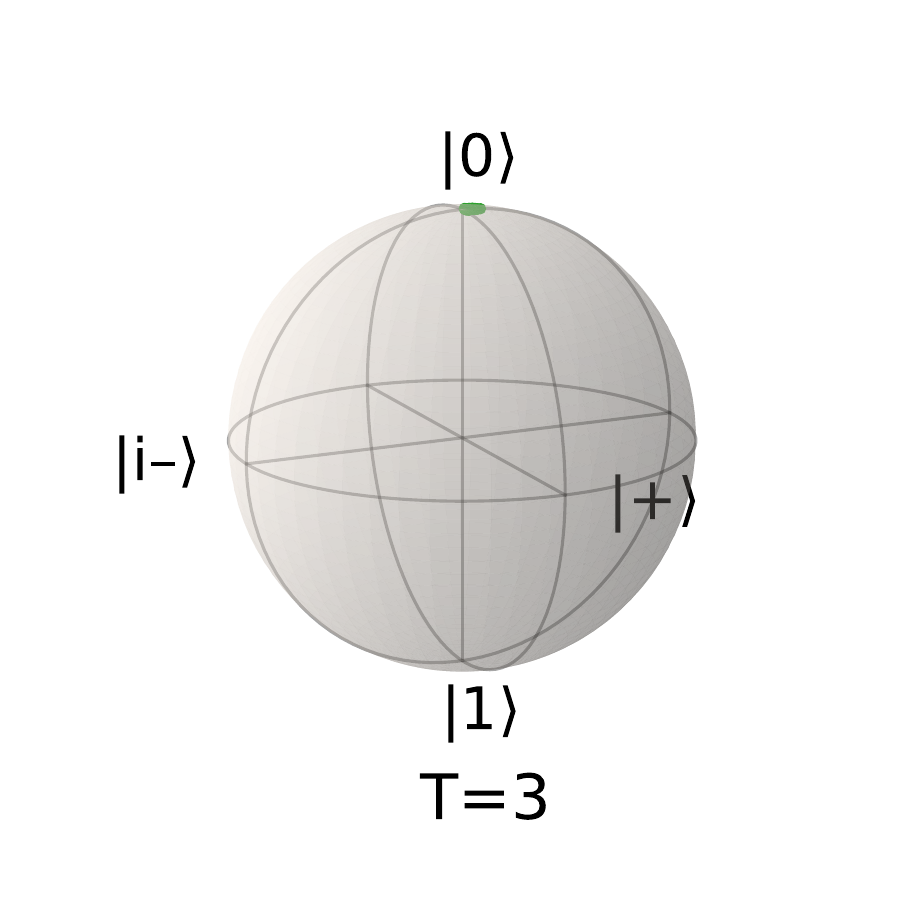}
\end{minipage}%
\begin{minipage}{.166\textwidth}
  \centering
  \includegraphics[width=1.3\linewidth]{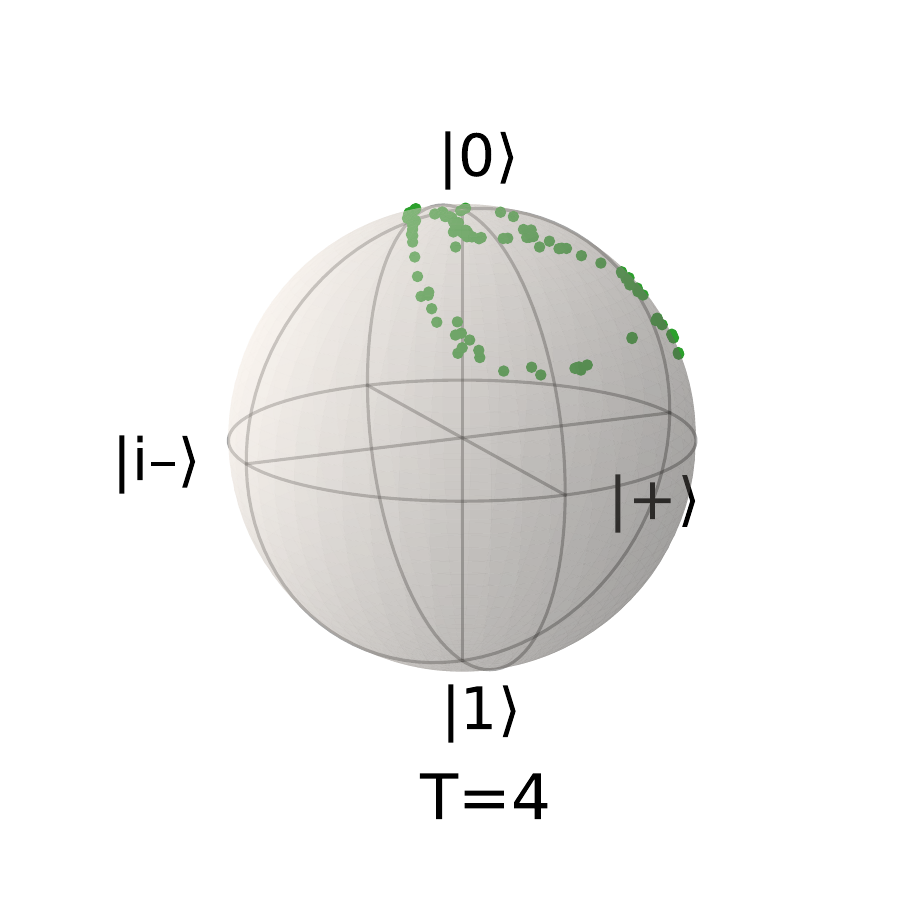}
\end{minipage}%
\begin{minipage}{.166\textwidth}
  \centering
  \includegraphics[width=1.3\linewidth]{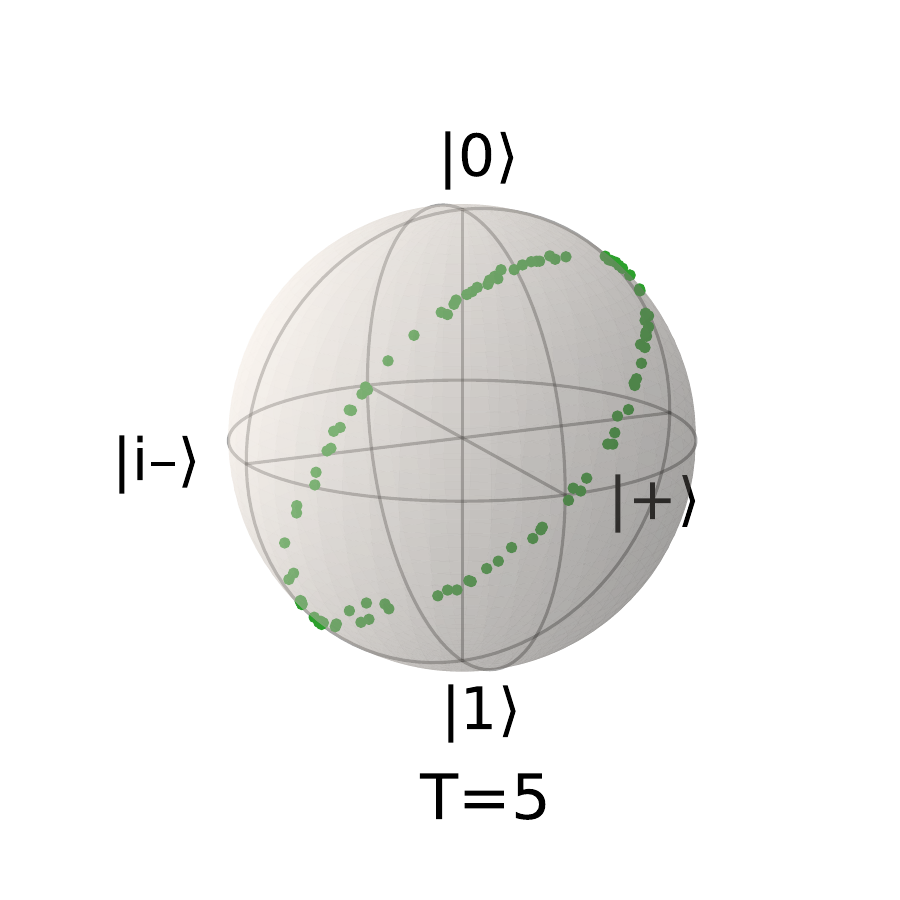}
\end{minipage}%
\caption{
The single qubit preparation results of 100 random samples.
From $T=0$ to $T=3$, the RNN prepares all mixed initial states to $|0\rangle$.
From $T=4$ to $T=5$, the RNN prepares the target ground state based on the input Hamiltonian.
}
\label{single_qubit}
\end{figure*}

For test our trained model, we select the test Hamiltonian as $H = \cos(\theta)\sigma_x + \sin(\theta)\sigma_y /\sqrt{2}+ \sin(\theta)  \sigma_z/\sqrt{2}$, where $\theta \in [0,2\pi)$ with uniform distribution (which is a subset of all training Hamiltonians).
Figure \ref{single_qubit} shows 100 samples of $\rho_{i}$ at each time steps.
The RNN takes the strategy to divide state preparation into two stages:
From $T = 0$ to $T = 3$, the RNN prepares the pure state $|0\rangle$ from whatever random initial states $\rho_0$.
From $T = 4$ to $T = 5$, the RNN rotates $|0\rangle$ to the target ground state specified by the given single-qubit Hamiltonian.
The measurement outcomes are effectively used by the RNN at all time steps so that it can learn the state of the system and accordingly propose the future quantum operation.
However, the information of the Hamiltonian is only effectively used in the last two steps, as the RNN aims to bring all initial states to $|0\rangle$ within the first four time steps.
\par
%
\subsection{Two-qubit State Preparation}
\label{sec:level3,2}
For the two-qubit preparation, we adopt $n=8$ time steps and the same quantum circuit structure as shown in Fig.~\ref{circuit}.
The initial states are pure states.
The randomly sampled two-qubit Hamiltonian takes the form of
\begin{equation}
\label{eq9}
    H = \sum_{ab} h_{ab}\sigma_{a}^1\sigma_{b}^2,
\end{equation}
where $h_{ab} \in \left[-1,1\right]$ and $a,b = x, y, z, I$.
\par
\begin{figure}[tb]
\includegraphics[width=1.0\linewidth]{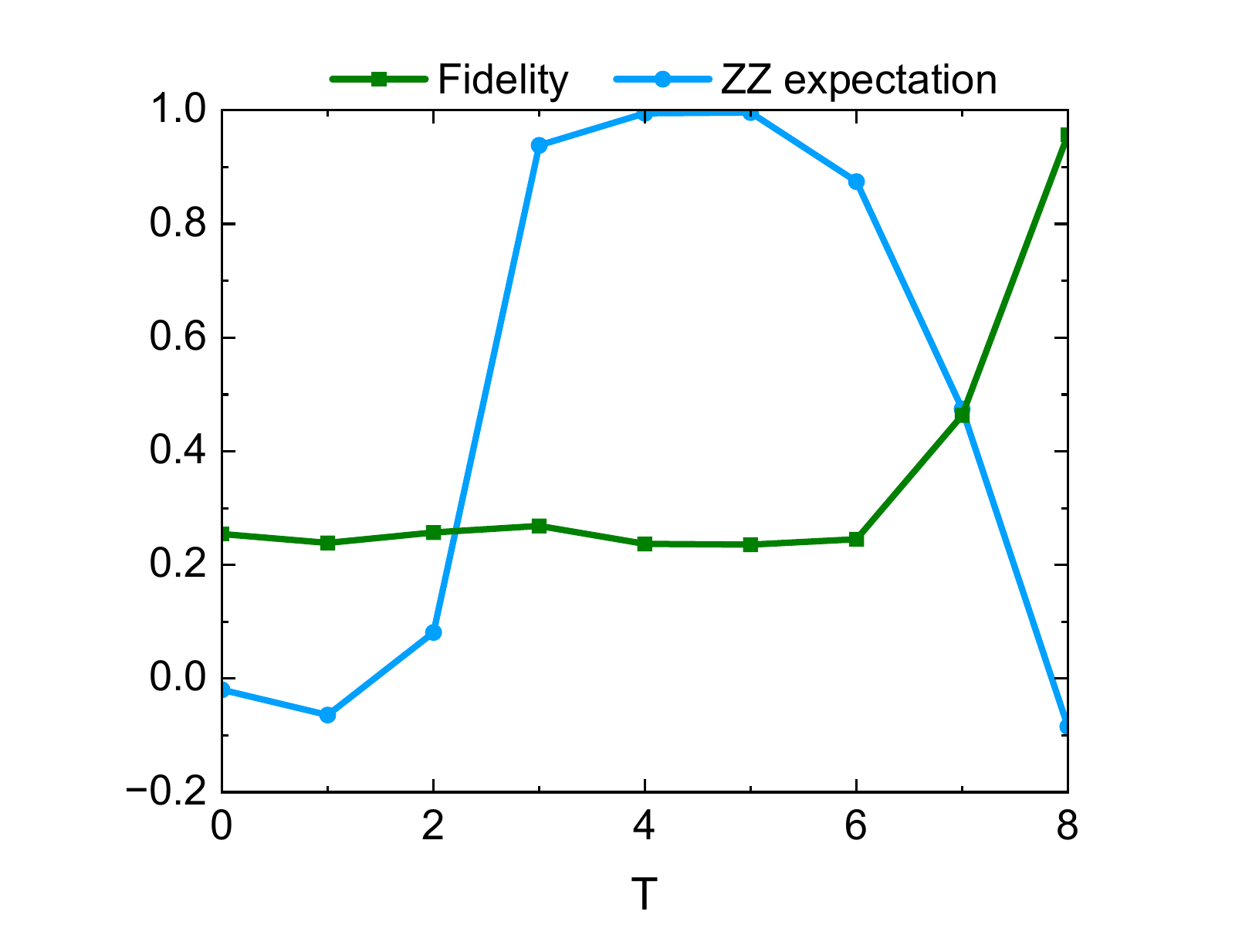}
\caption{
The average fidelity and expectation value of $\sigma^{1}_z \sigma^{2}_z$ of 100 samples at different time steps.
The ZZ expectation approaches one at $T= 4$ and $T=5$, indicating that the RNN first prepare all the initial states to $|00\rangle$, and subsequently to the ground state of the sampled Hamiltonian.
The fidelity is significantly increased only at the last two time steps, aligning with ZZ-expectation results.
}
\label{two_qubit}
\end{figure}

After training, we select the Hamiltonian of Eq.~\eqref{eq9} as the test Hamiltonian with 100 random samples.
Figure \ref{two_qubit} exhibits simulation results, where the blue line denotes the average fidelity, and the green line denotes the average expectation value of the Hamiltonian $\sigma_z \sigma_z$.
The results of ZZ-expectation values suggest that the RNN follows a strategy similar to the single-qubit case: to prepare from any random input state to $|00\rangle$ from $T=0$ to $T=5$, and then to prepare the target ground state from $|00\rangle$ from $T=6$ to $T=8$.
Consistent with the ZZ-expectation results, the fidelity significantly increases only in the last two time steps.
\par
\par
%
\subsection{Explore the Role of Ancilla Qubits}
\label{sec:level3,3}
Universally, there are three ways that ancilla qubits can be treated in each step of the quantum process: to be traced out, to be measured with feedback, and to be measured and without feedback.
Let $N^\text{anc}$ be the total number of ancilla qubits, which further patitions into $N_m^\text{anc}$ ancilla qubits for measurement-and-feedback and $N_t^\text{anc}$ ancilla qubits to be traced out. We would like to explore how the choice of these hyper-parameters $(N_m^\text{anc}, N_t^\text{anc})$ affects the performance of our model.
In the following experiement, we choose $N^\text{anc}=2,3,4$, with $N_m^\text{anc}$ ranging from $N_m^\text{anc} = 1$ to $N_m^\text{anc}$, and $N_t^\text{anc} = N^\text{anc} - N_m^\text{anc}$.
\par
We evaluate the performance of the RNN model by the  average fidelity between the model prepared state $\rho_T^\text{\text{sys}}$ and the target ground state $\rho_T^\text{min}$ over 100 random samples.
Because the optimization of the RNN fluctuates each time, each choice of the hyper-parameter $(N_m^\text{anc}, N_t^\text{anc})$ is trained three times to select the one with highest average fidelity for comparison.
\par
Table \ref{Fid} lists the average fidelity for each choice of $(N_m^\text{anc}, N_t^\text{anc})$. We observe the following features.
\begin{itemize}
    \item Increasing the total number of ancilla qubits, $N^\text{anc}$, improves accuracy. This is because additional quantum resources facilitate more effective ground state preparation.
    \item Accuracy can also be enhanced by allocating more ancilla qubits for measurement and feedback. A greater $N_m^\text{anc}$ indicates more measurement readout in each step, which provides the RNN with more information about the quantum system, enabling it to make more precise control decisions for ground state preparation task.
    \item However, using all ancilla qubits for measurement and feedback $N^\text{anc} = N^\text{anc}_{m}$ instead significantly degrades performance, results in the lowest accuracy. This occurs because projective measurement only increases the purity and reduces entropy of the quantum state. If all ancilla qubits are measured at every step, the system's entropy will decrease monotonically, preventing entropy-producing processes such as dissipation or decoherence and restricting the possible state preparation trajectories. By reserving at least one ancilla qubit to be traced out, the RNN can regulate system's entropy in both directions, enabling a more flexible and effective state preparation process.
\end{itemize}

\par
\begin{table}[tb]
  \centering
  \caption{
  The average fidelity of 100 samples for different choices of ancilla qubits in the case of two-qubit ground state preparation.
  }
  \label{Fid}
  \begin{ruledtabular}
    \begin{tabular}{@{\hskip 0.1in}c@{\hskip -0.5in}c@{\hskip 0.8in}}
                    &Fidelity  \\
\hline
$N = 2,\;\; N_m = 1$&0.908\\
$N = 2,\;\; N_m = 2$&0.853\\
\hline
$N = 3,\;\; N_m = 1$&0.913 \\
$N = 3,\;\; N_m = 2$&0.929 \\
$N = 3,\;\; N_m = 3$&0.879 \\
\hline
$N = 4,\;\; N_m = 1$&0.914 \\
$N = 4,\;\; N_m = 2$&0.940 \\
$N = 4,\;\; N_m = 3$&0.953 \\
$N = 4,\;\; N_m = 4$&0.888 
    \end{tabular}
  \end{ruledtabular}
\end{table}
%
\section{Summary}
\label{Sect:V}
In this paper, we proposed an RNN-based approach for ground state preparation, utilizing mid-circuit measurement and feedback in a multi-step process to extract state information. 
Unlike previous works that applied machine learning to search for the ground state on a case-by-case basis, our RNN dynamically controls quantum circuits based on measurement outcomes and feedback. 
Additionally, by training on diverse Hamiltonians, the RNN learns to infer ground state preparation protocols that adapt to a wide range of Hamiltonians. This enables the algorithm to be trained once and applied across various scenarios as a foundational model.
\par
Interestingly, we observed that the RNN does not prepare the target ground state by approaching to the target state gradually at each time step.
Instead, it first drives all initial states toward an intermediate state before transitioning to the target state. 
Additionally, we found that utilizing more ancilla qubits for measurement and feedback improves the RNN's control protocols. However, at least one ancilla qubit should be reserved for tracing out to introduce dissipation, which is essential for effective state preparation.
\par
Our work is applicable to experimental implementations with certain adjustments to account for differences between classical and quantum hardware. Since quantum circuits cannot be optimized using gradient-based methods on real quantum hardware, non-gradient-based optimization techniques \cite{TILLY20221} can aid in optimizing the RNN parameters. 
Currently, our current model does not incorporate quantum noise, which could impact accuracy particularly as the system size increases. 
Therefore, an important direction for future work is to integrate quantum noise into our model.
%
%
\begin{acknowledgments}
We acknowledge the inspiring discussion with Jing-Ze Zhuang.  Y.Z.Y. is supported by the NSF Grant No. DMR-2238360. We acknowledge OpenAI ChatGPT 4o for providing editing and language suggestings in writing this paper.
\end{acknowledgments}
%
\bibliography{ref}
%
\end{document}